\def \MESS #1 #2 {{\em The Messenger\/} {\bf #1}, #2}
\def \ASTRNACH #1 #2 {{\em Astron. Nach.\/} {\bf #1}, #2}
\def \AAP #1 #2 {{\em Astron. Astrophys.\/} {\bf #1}, #2}
\def \AAL #1 #2 {{\em Astron. Astrophys. Lett.\/} {\bf #1}, L#2}
\def \AAR #1 #2 {{\em Astron. Astrophys. Rev.\/} {\bf #1}, #2}
\def \AAS #1 #2 {{\em Astron. Astrophys. Suppl. Ser.\/} {\bf #1}, #2}
\def \AJ #1 #2 {{\em Astron. J.\/} {\bf #1}, #2}
\def \ANNREV #1 #2 {{\em Ann. Rev. Astron. Astrophys.\/} {\bf #1}, #2}
\def \APJ #1 #2 {{\em Astrophys. J.\/} {\bf #1}, #2}
\def \APJL #1 #2 {{\em Astrophys. J. Lett.\/} {\bf #1}, L#2}
\def \APJS #1 #2 {{\em Astrophys. J. Suppl.\/} {\bf #1}, #2}
\def \APSS #1 #2 {{\em Astrophys. Space Sci.\/} {\bf #1}, #2}
\def \ASR #1 #2 {{\em Adv. Space Res.\/} {\bf #1}, #2}
\def \BAIC #1 #2 {{\em Bull. Astron. Inst. Czechosl.\/} {\bf #1}, #2}
\def \JSQRT #1 #2 {{\em J. Quant. Spectrosc. Radiat. Transfer\/} {\bf #1}, #2}
\def \MN #1 #2 {{\em Mon. Not. R. Astr. Soc.\/} {\bf #1}, #2}
\def \MEM #1 #2 {{\em Mem. R. Astr. Soc.\/} {\bf #1}, #2}
\def \PLR #1 #2 {{\em Phys. Lett. Rev.\/} {\bf #1}, #2}
\def \PASJ #1 #2 {{\em Publ. Astron. Soc. Japan\/} {\bf #1}, #2}
\def \PASP #1 #2 {{\em Publ. Astr. Soc. Pacific\/} {\bf #1}, #2}
\def \NAT #1 #2 {{\em Nature\/} {\bf #1}, #2}

\documentstyle[twoside]{memsait}
\input psfig.tex
\begin{opening}
\title {A NEW CLASS OF X--RAY PULSARS ?}
\author{L. STELLA$^{1,2,5}$, S. MEREGHETTI$^3$, G.L. ISRAEL$^{4,5}$}
\institute{$^1$ Osservatorio Astronomico di Brera, Via E, Bianchi, 46, I--22055 
Merate (Como), Italy.\\
$^2$ Now at: Osservatorio Astronomico di Roma, Via dell'Osservatorio 2, I--00040 
Monteporzio \\ ~~(Roma), Italy.\\
$^3$ Istituto di Fisica Cosmica, CNR, Via Bassini 15, I--20133 Milano, Italy.\\
$^4$ International School for Advanced Studies (SISSA--ISAS), Via Beirut, 3, 
I--34014, Trieste, \\ ~~Italy\\
$^5$ Affiliated to I.C.R.A.}
\date{} % DO NOT INSERT ANY DATE HERE !!!

\end{opening}

\begin{document}
\oddpagefooter{}{}{} % LEAVE AS IT IS !
\evenpagefooter{}{}{} % LEAVE AS IT IS !
\ 
\bigskip

\begin{abstract} 

While the distribution of spin periods of High Mass X--ray Binaries 
(HMXBs) spans more than four orders of magnitude (69~ms -- 25~min), the few 
known X--ray pulsars accreting from very low mass companions 
($<1$~M$_{\odot}$) have very similar periods between 5.4 and 8.7~s. 
These pulsars 
display also several other similarities and they are probably
members of a subclass of Low Mass X--ray Binaries (LMXBs) with 
similar  magnetic field (a few $10^{11}~$G), companion stars and, 
possibly,  
evolutionary histories. If they are rotating at, or close to, the 
equilibrium period, their properties are consistent with luminosities 
of the order of a few $10^{35}$~erg~s$^{-1}$.  These pulsars might 
represent 
the closest members of a subclass of LMXBs characterized by lower 
luminosities, higher magnetic fields and smaller ages than 
non--pulsating LMXBs. 
\end{abstract}

\section{Introduction}

X--ray pulsations with periods ranging from 
$\sim 0.069$ to $\sim 1450$~s are present in a large number (about 45)
of X--ray binaries. This signal originates from the beamed radiation 
which is produced close to
the magnetic poles of a young accreting neutron star with a surface field of 
$\sim 10^{12}-10^{-13}$~Gauss. Due to the misalignement of the magnetic and
rotation axes, the neutron star rotation modulates  
the X--ray intensity observed at the earth in a light--house fashion. 
Period (or phase) changes, introduced by the binary motion, allow to measure 
some of the orbital parameters of these systems.
Together with the duration of the X--ray eclipse (which is observed in several 
systems) and the Doppler velocity and photometric modulations of the optical
companion, these measurements provide the absolute orbital solution and the masses 
of the two components. 

Most of these  pulsars (over 35) are found 
in high mass X--ray binaries (HMXRB), containing an early type (OB) star 
with a mass of $>5$~M$_{\odot}$. Mass transfer 
usually takes place because part of the intense
stellar wind  emitted by the OB star is captured by the gravitational field
of the collapsed  object. In some cases the OB star fills up its
Roche lobe and mass transfer, towards the neutron star, takes place also
through the first Lagrangian point, leading to the formation of an accretion
disk around the neutron star. 
Secular spin period changes arise because of the torque exerted on the
neutron star magnetosphere by the accreting matter 
(Henrichs 1983 and references therein). 
Disk--fed systems are characterised by a pronounced spin--up with a timescale
that can be as short as $\sim 100$~yr. Alternating spin--up and spin--down 
intervals are instead frequent among  wind fed--systems;
these likely result from the variable wind characteristics of the mass donor
early type star (especially if a Be star), which can cause the angular
momentum of the captured material to reverse its sign relative to the
neutron star.  Even in these cases, however, an average spin--up trend 
is ensued.  

%The energy production budget of X--ray pulsars in High Mass X--ray 
%Binaries (HMXBs) is often 
%dominated by the optical luminosity of the OB star, with the X--ray
%luminosity emitted in  the vicinity of the collapsed object providing 
%only a small  perturbation ($L_x/L_{opt} $$\sim$$ 10^{-3}-10$).
%Correspondingly, the optical spectra are 
%stellar--like\cite{RaJo}.
%%%(Rappaport and Joss 1983, and references therein)

The spectra of X--ray pulsars in HMXBs are quite hard. They are usually 
characterised in terms of a power--law extending up to energies of several
tens of keV, followed by a steep nearly exponential decay, beyond which 
cyclotron absorption features are sometimes detected. 

The number of X--ray pulsars that are not in HMXBs is very  small, but recent
observations have virtually doubled this sample. This  has also resulted from
the finding that a few optically unidentified sources  have X--ray to optical
flux ratios incompatible with the presence of  massive companions. 
Several of these systems have low mass donor stars and, therefore, are 
Low Mass X--ray Binaries (LMXBs). 
Two of the  LMXBs pulsars,  Her~X--1, which has a comparatively massive 
companion of $\sim 2$~M$_{\odot}$, and GX~1+4, 
whose companion is an M giant, are  
peculiar systems (see, e.g., Rappaport \& Joss 1983, Nagase 1989).  The 
remaining five LMXBs pulsars (4U~0142+61, 1E~1048.1--5937, 4U~1626--67, 
RX~J1838.4--0301 and 1E~2259+589) have very similar spin periods in the 
5 to 9~s range (see  Fig.~1). This narrow period distribution is 
remarkable, when one considers that the X--ray pulsars in HMXBs have 
spin periods ranging from 69~ms (A~0538--66, Skinner et al. 1982) to 
25~min (RX~J0146.9+6121, Mereghetti, Stella \& De Nile 1993). We 
proposed that these LMXB pulsars belong to a homogeneous class of sources 
and suggest a possible explanation for their narrow spin period distribution
(Mereghetti \& Stella 1995). 
\begin{figure}[bht]
\centerline{\psfig{figure=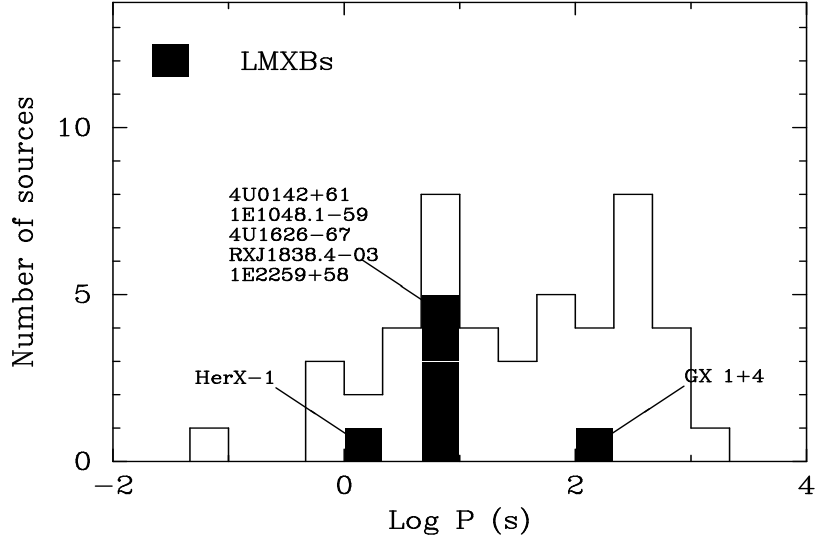,width=14cm,height=8.5cm} }
\caption[h]{The observed period distribution of X--ray pulsars. The black 
squares indicate the X--ray pulsars that are not HMXBs.}
\end{figure}

\section{Observational characteristics of the Very Low Mass X--ray Binary 
Pulsars} 

\noindent{a) 4U~0142+61}  

The properties of 4U~0142+61 (White et al. 1987) remained puzzling for a 
long time, owing to confusion problems with a nearby pulsating and 
transient Be/neutron star system 
(Motch et al. 
1991, Mereghetti, Stella \& De Nile 1993). Despite the small error box 
(5" radius), no optical counterpart has yet been identified, down to 
V$>24$  (Steinle et al. 1987), 
thus excluding the presence of a massive companion.   
Using data from the EXOSAT archive, Israel, Mereghetti \& Stella (1994) 
discovered periodic pulsations at 8.7 s, which were later confirmed 
with ROSAT (Hellier 1994). No delays in the pulse arrival times caused by 
orbital motion were found, with upper limits on $a_x\sin i$  of 
about $\sim 0.37$~lt--s for 
orbital periods between 7~min and 12~hr (Israel, Mereghetti \& Stella 1994). The EXOSAT 
and ROSAT period measurements, obtained in 1984 and 1993, provide a 
spin--down rate of $\sim 7\times10^{-5}$~s~yr$^{-1}$.
The 2--10~keV spectrum, a power law with photon index of $\sim 4$, is 
extremely soft (White et al. 1987) and led to the initial classification of 
this source as a possible black hole candidate. The X--ray luminosity of 
4U~0142+61 did not show large secular variations around an average value of 
$\sim~2.5 \times 10^{35}$~erg~s$^{-1}$ (assuming a distance of 2~kpc). 
%A local ($<1$~kpc) molecular 
%cloud observed in this direction (Leisawitz et al. 1989) can easily account 
%for the column density measured in the X--ray spectrum, 
%$N_H \sim 1.5\times 10^{22}$~cm$^{-2}$, 
%without the need of invoking a much larger distance.  

\noindent {b) 1E~1048.1--5937}  

This source was serendipitously discovered with the Einstein Observatory 
in 1979, and found to pulsate at 6.44~s (Seward, Charles \& Smale 1986). 
The brightest candidate counterparts in the small error box  
have V$>20$ (Mereghetti, Caraveo \& 
Bignami 1992)  indicating that also 1E~1048.1--5937 is a LMXB.
1E~1048.1--5937 was repeatedly observed with ROSAT  in 1992 and 1993.
While all the previous observations  with EXOSAT and GINGA (Corbet \& Day 1990)
were consistent with a constant spin-down at a rate 
of $\sim 5 \times 10^{-4}$~s~yr$^{-1}$,
the ROSAT data (Mereghetti 1996)
indicate a doubling of the spin-down rate (Fig.~2) .
\begin{figure}[tbh]
\centerline{\psfig{figure=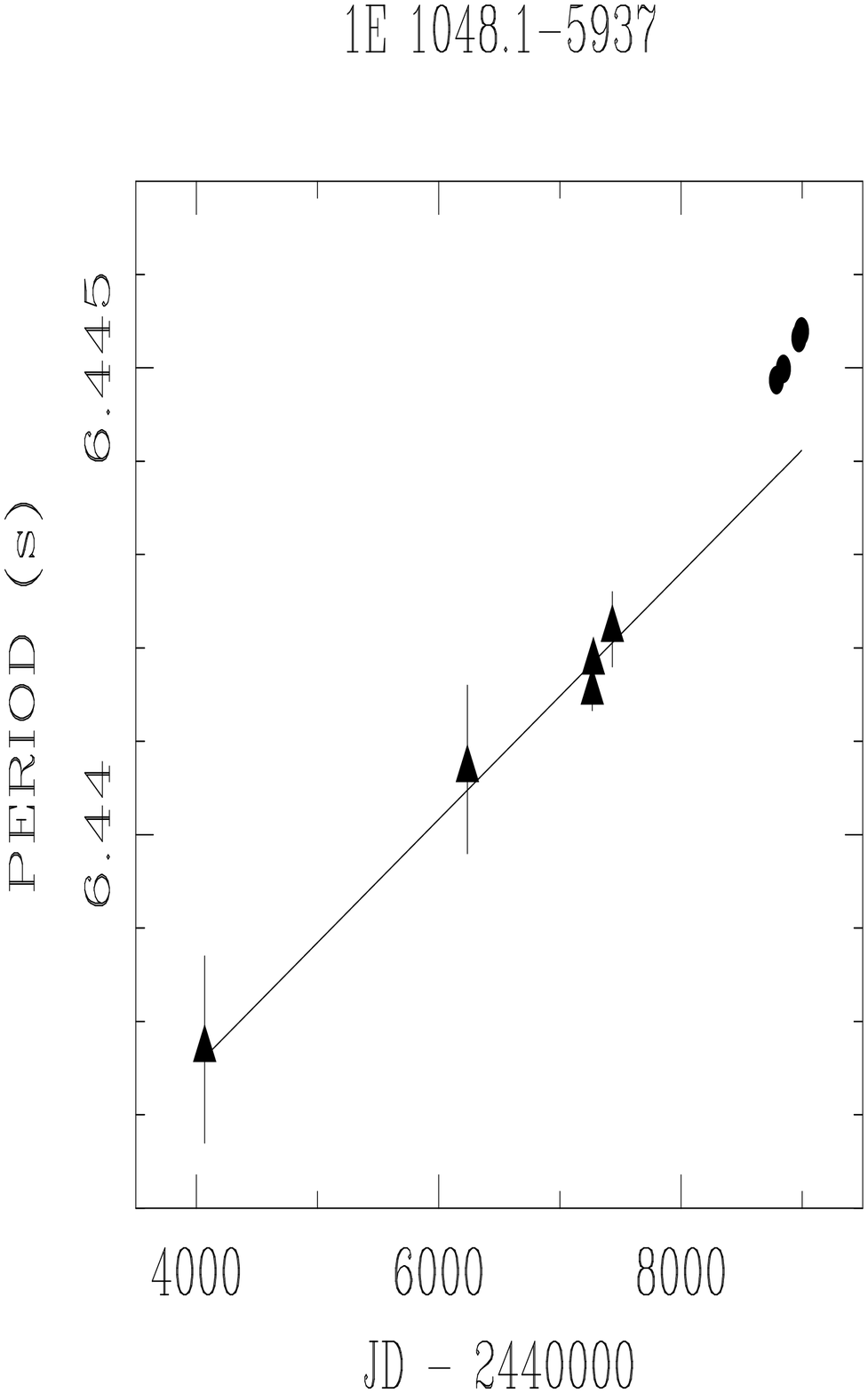,width=14cm,height=8.5cm} }
\caption[h]{Spin period evolution of 1E~1048.1--5937 (from Mereghetti 1995).}
\end{figure}
The power law photon index derived with EXOSAT (2.3, Seward et 
al. 1986) implies also for this source a spectrum somewhat softer than the 
``canonical" spectrum of X--ray pulsars. The high column density suggests 
that 1E~1048.1--5937 lies behind the Carina nebula, i.e. at more than 
2.8~kpc. The luminosity corresponding to this distance is 
$\sim 2 \times 10^{34}$~erg~s$^{-1}$. 
 
\noindent{c) 4U~1626--67}  

This  LMXBs  is optically identified with the V$\simeq$ 18.5, blue star KZ TrA
(Mc Clintock et al. 1977). In addition to
the X--ray  periodicity at 7.7~s (Rappaport et al. 1977), a pulsation at a
slightly lower  frequency is present in the optical band (Middleditch et al.
1981). This is  probably due to reprocessing of the X--ray pulses occurring
near the  companion star, and the difference of the two periodicities can be 
explained with an orbital period of 41.4~min. The high X--ray to optical flux 
ratio, as well as the very strong limits on the optical mass function 
($a_x\sin i$ $ <$ $ 0.013$~lt--s  for 10~min~$< P_{orb}<$~10~hr, Levine et al.
1988), clearly indicate  that 4U~1626--67 is a LMXB. The period measurements
obtained before  1990 were consistent with a constant spin--up rate of
$-1.6 \times 10^{-3}$~s~yr$^{-1}$ (Nagase  1989),
but in  1991 the period derivative changed sign 
(Lutovinov et al. 1994, Bildsten et al. 1994).  
On the basis of the
luminosity required to explain the spin--up  torque 
($L_x\sim2\times 10^{36}$$ - $$10^{37}~erg~s^{-1}$) Levine et al. (1988) estimated a
distance  of 3 to 6~kpc. With the exception of quasi--periodic
flares with a characteristic timescale  of $\sim 1000$~s, little intensity
variations are present in 4U~1626--67. The pulse  averaged spectrum, a flat
power law (photon index $\sim 0.4$) followed by an  exponential cut--off at
$\sim 20$~keV (Pravdo et al. 1979), is  similar to that of  HMXB pulsars
(White, Swank \& Holt 1983).  
A recent ASCA observation revealed the presence of spectral features 
around   1 keV, interpreted as emission lines of hydrogen--like Neon
(Angelini et al. 1995). This indicates an overabundance of Ne 
in 4U~1626--67, which might give interesting constraints
on the nature and evolution of its companion star.

\noindent{d) RX~J1838.4--0301}

This pulsar, recently discovered with ROSAT, is embedded in a region of 
diffuse X--ray and radio emission, which is interpreted as a supernova 
remnant $\sim 32000$ years old at a distance of $\sim 4$~kpc (Schwentker
1994). The  brightest optical object in its 10" radius error box has V=14.
The spectrum in the 0.1--2.4~keV band, and therefore the unabsorbed flux, are
not well  constrained by the data (Schwentker 1994), but there
is  evidence that also this source is quite soft (best fit power law photon 
index of $\sim 3$). Also RX~J1838.4--0301 is likely a LMXB.  

\noindent{e) 1E~2259+586}  

The source 1E~2259+586 was discovered with the Einstein Observatory at 
the center of the X--ray and radio supernova remnant G109.1--1.0 (Fahlman 
\& Gregory  1981). Extensive searches for optical, IR and radio counterparts 
were carried out without success (Fahlman et al. 1982; Coe \& Jones 1992;  
Coe, Jones \& Letho 1994), but they definitely exclude the presence of a 
massive companion (Davies \& Coe 1991). 
The spin period of 1E~2259+586 has  been increasing   
at  $\sim 2\times 10^{-5}$~s~yr$^{-1}$ until 1992 (Koyama et al. 1989, 
Iwasawa, Koyama \& Halpern 
1992).  Recent ROSAT data revealed the first spin--up episode for this source.
A detailed analysis of all the period measurement over the last 15 years
(Baykal \& Swank 1996) showed that  1E~2259+586 undergoes random angular
velocity variations similar to those observed in other accreting 
binary neutron stars (Fig.~3).
\begin{figure}[bht]
\centerline{\psfig{figure=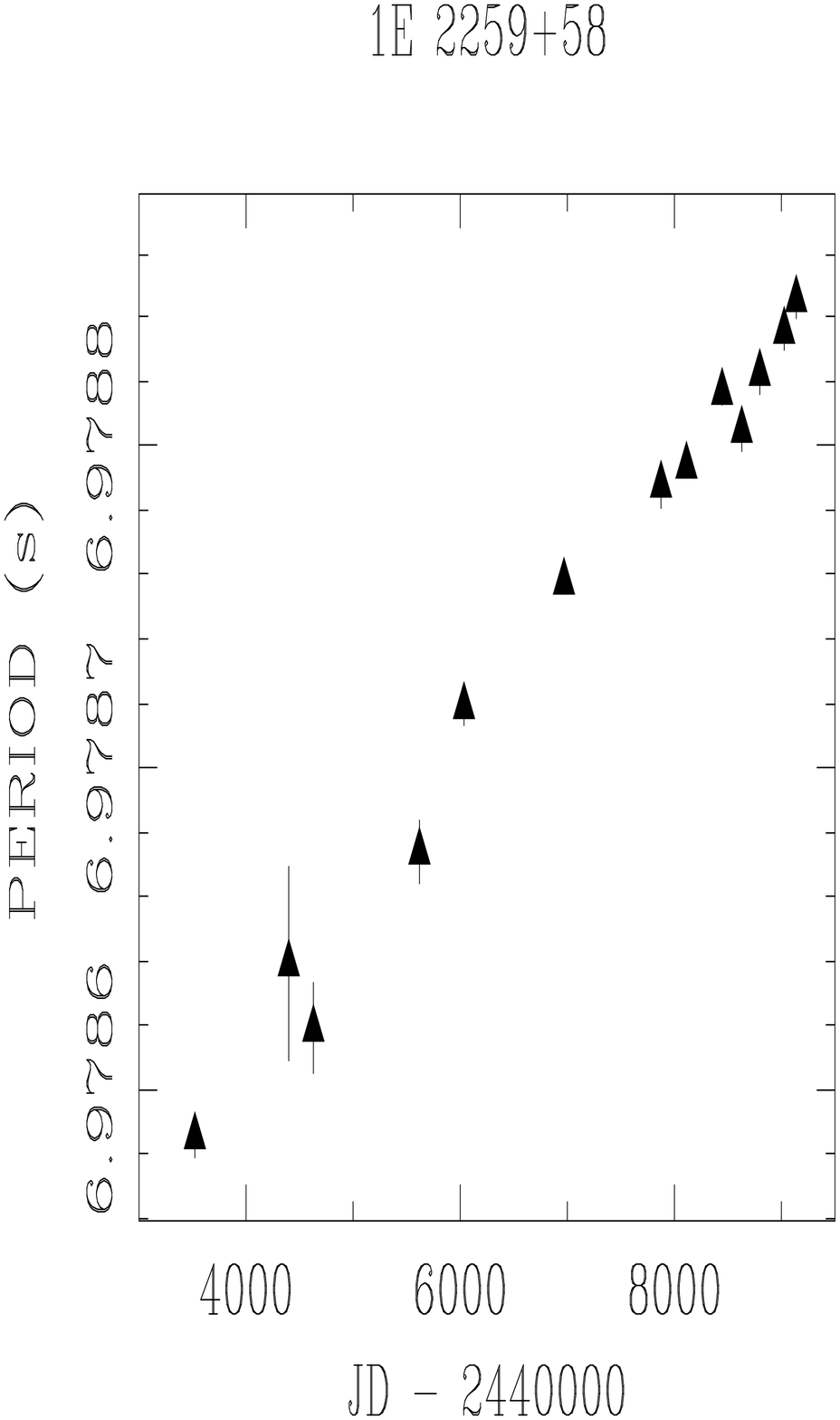,width=14cm,height=8.cm} }
\caption[h]{Spin period evolution of 1E 2259+58.}
\end{figure}
The upper limit on $a_x\sin i$  is of 
0.08~lt--s for 10$^3$~s $< P_{orb}<$ 10$^4$~s   (Koyama  et al. 1989).
%The spectral analysis of 1E~2259+586 with non--imaging instruments is  
%complicated by the presence of the surrounding diffuse emission due to 
%the supernova remnant and the galactic ridge emission (Koyama et al. 
%1989). The results for the pulsed flux, less sensitive to this problem, 
%indicate a very steep power law spectrum with photon index $\sim$4--5 (Morini 
%et al. 1988, Hanson et al. 1988, Iwasawa et al. 1992). 
1E~2259+586 has a very     soft 
spectrum, as     recently confirmed by BBXRT and ASCA observations 
(Corbet et al. 1995). Also in this source, no long term variability greater 
than a factor $\sim$2 has been reported. A distance of 3.6$\pm$0.4 kpc for the 
supernova remnant G109.1--1.0 was estimated by using the classical 
relation between distance and radio surface brightness (Gregory \& 
Fahlman 1980). Based on improved radio observations, Hughes et al. 
(1984) have subsequently derived a value of 5.6 kpc. For this distance the 
average X--ray  luminosity of 1E~2259+586 is $\sim$2 $\times 10^{35}$ erg s
$^{-1}$. 

\section{A new class of X--ray Pulsars ? } 

The main properties of the sources considered above are summarized in  
Table 1.  Their most important similarities are the following.
\vspace{-6mm}
\begin{table}[h]
\begin{center}
%Table 1
\caption{Properties of the VLMXB pulsars}
\begin{tabular}{lccclcr}
\hline \hline \\
Source&Period&Spin--down&X--ray Flux&Pow.--Law& L$_{eq}$&d$_{eq}$\\
& (s)  &(s yr$^{-1}$)&(erg cm$^{-2}$ s$^{-1}$)& Ph.In.&(erg s$^{-1}$)& (kpc)\\
 \\ \hline \\
4U 0142+61&8.69&7.2$\times$ 10$^{-5}$&5.5$\times$ 10$^{-10}$&	
4	&1.4$\times$ 10$^{35}$&1.5\\
1E 1048.1--59&
	6.44&	4.6$\times$ 10$^{-4}$&	2.2 $\times$10$^{-11}$&2.3&
        2.8$\times$ 10$^{35}$&	10.6\\
4U 1626--67&
	7.66&	1.4$\times$ 10$^{-3}$&	6.0 $\times$10$^{-10}$  a&0.4&
        1.9$\times$ 10$^{35}$&	1.7\\
RX J1838.4--03&
	5.45&	...&	5$\times$10$^{-12}$--4$\times$10$^{-9}$&	3&
        4.1$\times$ 10$^{35}$&	27--1\\
1E 2259+58&
	6.98&	2.3$\times$ 10$^{-5}$&	5.3$\times$ 10$^{-11}$&	4--5&
        2.3$\times$ 10$^{35}$&	6.2\\ \\
        \hline \\
\end{tabular}
\end{center}
NOTE - $L_{eq}$ and $d_{eq}$ have been computed assuming $B=5$$~\times$ 10$^{11}$ 
G for all the sources.

$^a$  flux during spin--down; when 4U 1626--67 was spinning--up at --1.6 
$\times$10$^{-3}$ s yr$^{-1}$ its flux was $\sim$4 times higher.
\end{table}
(a) When compared to the other X--ray pulsars, the most striking property 
of  these objects is their narrow spin period distribution (Fig.~1).  A 
Kolmogorov--Smirnov test yields a probability of 4 $\times$10$^{-3}$ that the 
period distribution of these five sources and that of the HMXBs pulsars are 
drawn from the same parent distribution.
(b) With the  exception of 4U~1626--67, their spectra are much softer than 
those of the other X--ray pulsars (White, Swank \& Holt 1983).  Figure~4 shows a 
comparison of the EXOSAT spectra of Vela~X--1 and 4U~0142+61. 
\begin{figure}[thb]
\centerline{\psfig{figure=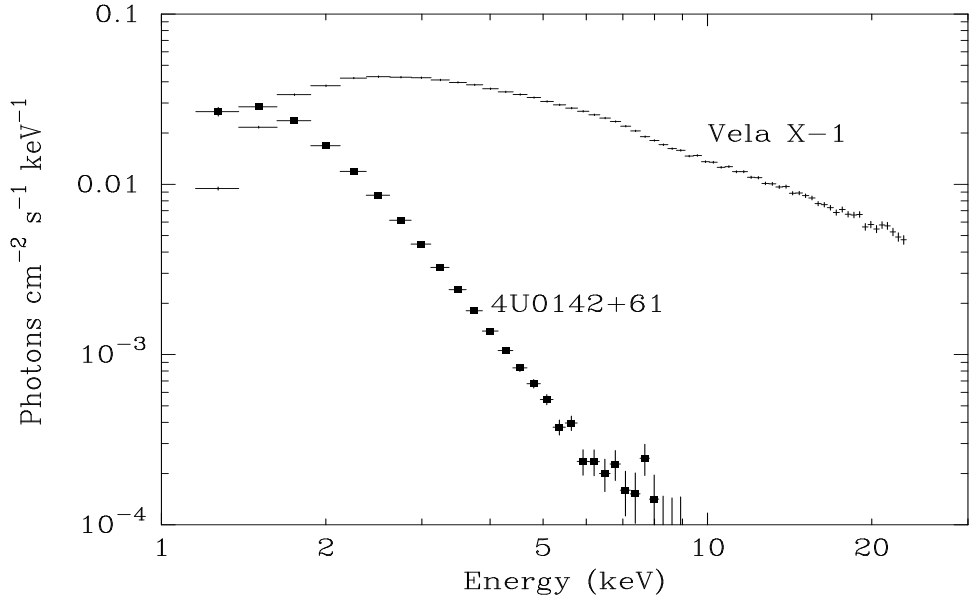,width=14cm,height=8.5cm} }
\caption[h]{A comparison of the EXOSAT ME spectra of Vela X--1 and 4U 0142+61.}
\end{figure} 
The steep power law spectra of 4U~0142+61 and 1E~2259+586 are even softer
than those of  most LMXBs (White, Stella \& Parmar  1988). This is also 
apparent from the position occupied by these source in the 
X--ray colour--colour diagram of Fig.~5.
\begin{figure}[bht]
\centerline{\psfig{figure=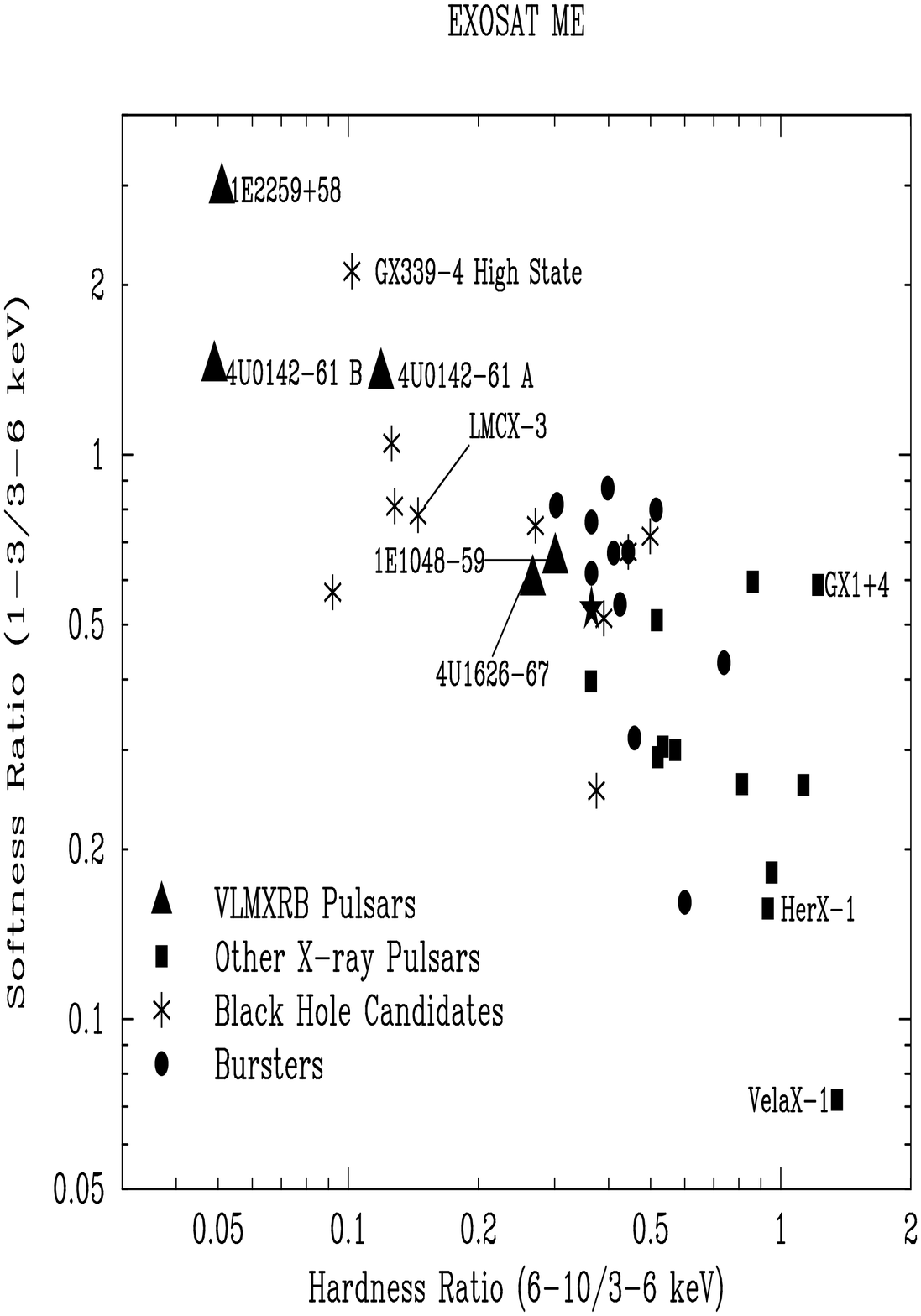,width=14cm,height=8.5cm} }
\caption[h]{X--ray colour--colour diagram for a sample of X--ray binaries.}
\end{figure}
(c) These sources are probably less luminous than most persistent LMXBs. 
Indeed, for 1E~2259+586, 1E~1048.1--5937 and 4U~0142+61 firm upper 
limits of a few 10$^{36}$ erg s$^{-1}$ can be derived by requiring that they 
lie within the Galaxy. The best estimates for the distances of 4U~1626--67 and 
RX~J1838.4--0301 also imply luminosities of this order of magnitude or 
smaller.
(d) Their flux appears to be relatively constant on  timescales  from 
months to years. This is unlike most of the other X--ray pulsars (including 
Her~X--1 and GX~1+4) for which large flux variations (encompassing 
transient activity)  have been observed.    
(e) Two of these sources are likely associated to supernova remnants. 
Another source, 1E 1048.1--5937, lies in the direction of the Carina Nebula, 
a complex region of radio, optical and X--ray diffuse emission, clearly 
associated with recent star formation activity. 

Based on these similarities, Mereghetti \& Stella (1995) proposed that these pulsars  
belong to a homogeneous subclass of LMXBs characterized by lower  
luminosities  ($10^{35}-10^{36}$~erg~s$^{-1}$) 
and higher magnetic fields ($B\sim10^{11}$~G)
than classical, non-pulsating LMXBs.

\section{Models} 

Models that have been proposed through the years for some of the X--ray 
pulsars of the new class differ with respect to the energy production 
mechanism as well as binary versus single nature of the magnetic degenerate
star. Among models based on rotational energy dissipation, simple radio
pulsar  models can be ruled out  based on the fact that measured spin--down rates
correspond to a rotational energy loss by a magnetic neutron star which is 
more than two orders of magnitude lower than the inferred X--ray
luminosities. Carlini and Treves (1989) suggest 
a modified radio pulsar scenario in which 
the emission beam of a freely precessing neutron star sweeps the
earth from  different angles, depending on the precession phase. 
In their model, the observed
X--ray periodicity   reflects the precession period 
of a weakly magnetized  (B $\sim 10^{10}$~G), neutron star with mass 0.3 
$M_{\odot}$ and with a spin period of only a few milliseconds.
%Carlini \& Treves (1989) show that for a spin period of
%a few milliseconds, an oblateness of a few percent and a magnetic
%field of $\sim 10^{10}$~G, the X--ray period and period derivative of
%1E~2259+586 ???check??? are reproduced, while the neutron star rotational
%energy loss is somewhat  higher than the measured X--ray luminosity. 
Morini et al. (1988) and Paczynski (1990) propose instead a white dwarf equivalent of the
standard  radio pulsar model. Due to the factor of $\sim 10^5$ larger moment
of inertia, the white dwarf rotational energy loss implied by the measured 
$\dot P$ is considerably larger than the measured X--ray luminosity.
Within this framework, the spin--down rate changes of 1E~2259+586 observed by
Iwasawa, Koyama \& Halpern (1992) have been interpreted by Usov (1994) in terms of 
white dwarf glitches.  
In a different vein, Thompson \& Duncan (1993) suggest that 1E~2559+586 
spins down like a standard radio pulsar, while the emitted 
radiation results from the gradual dissipation of the intense 
magnetic field ($>10^{14}$~G) of the neutron star. 
 
All the models outlined above are virtually ruled out by two recent
studies. Baykal \& Swank (1995) show that the spin period history of
1E~2259+586 consists of short term spin--up episodes superposed on a 
secular spin--down and that its fluctuation level is similar to that
of a number of accreting X--ray pulsars in HMXBs. More crucially 
Mereghetti (1996) revealed an increased spin--down episode from 
1E1048.1--5937, which is clearly incompatible with the spin--down 
rate expected in the unconventional applications of the radio pulsar 
model described above. 

Models based on matter accretion onto a magnetic rotating neutron star
are  clearly favored. These models, in turn, envisage both possibilities 
that the neutron star is isolated or in a binary system. 
Israel, Mereghetti \& Stella (1994) and Corbet et al. (1995) suggest that the
neutron stars in 4U~0142+614  and 1E~2259+586, respectively,  
might be accreting 
matter from a dense region of a molecular cloud. The problem with this 
scenario is that the highest expected mass capture rates are in the 
$10^{13}$g~s$^{-1}$ range, therefore giving rise to an accretion 
luminosity of $\sim 10^{33}$~erg~s$^{-1}$. This is well below the inferred 
X--ray luminosities. 

Based on the galactic distribution of four of the X--ray pulsars in the
sample (4U1626--67 is excluded because of its binary nature), 
van Paradijs et al. (1995) propose that they are the result of the evolution of 
a neutron star spiralling in a massive star, after the so--called
Thorne--Zytkov stage. Therefore, these sources should consist of 
isolated neutron stars accreting matter from a residual disk. 
While clearly viable, this model faces difficulties with the 
stability of such a self--gravitating disk (van Paradijs et al. 
predict a disk mass in the $10^{-3}-1$~M$_{\odot}$ range). 

Mass accretion
from a  companion star is the simplest explanation to account for the
observed X--ray emission.  In this framework, the peculiar period distribution 
of these sources can   
be explained by assuming that these neutron stars are
rotating close to their equilibrium period 
(see Mereghetti \& Stella 1995 for details). 
Due to the high angular momentum content,  mass transfer in LMXBs is 
mediated by an accretion disk, resulting in a  secular spin--up, 
unless the corotation radius is close to the size of the 
neutron star magnetosphere. In this ``equilibrium rotator" regime (cf. 
Ghosh \& Lamb 1979) a spin--down may result from the torques exerted 
by the magnetic field lines threading the accretion disk. The equilibrium 
rotator condition is given by : 
$$
  L_{eq} \sim  8.6 \times 10^{35} (B/10^{11} G)^2  P^{-7/3}   erg~ s^{-1} ,
$$
where $L_{eq}$ is the accretion luminosity at equilibrium and B the surface  
magnetic field of the neutron star (see, e.g., Henrichs 1983;  we   use
a neutron star mass of $1.4 ~ M_{\odot}$ and radius of
10$^6$ cm).  The measured spin--down rate of three of these sources  testifies 
that the neutron star is close to equilibrium. We assume that also 
RX~J1838.4--0301 and 4U~1626--67 are (nearly) equlibrium  rotators. In the 
case of the latter source this is supported by the recent reversal from 
spin--up to spin--down.
A measurement of the magnetic field strength based on cyclotron features 
is available only for 1E~2259+586, giving $B\sim 5 \times 
10^{11}$ G 
(Iwasawa, Koyama \& Halpern 1992, see, however, Corbet et al. 1995 for a different 
interpretation). The magnetic field of the other systems can be constrained 
by using the properties of their X--ray continuum. The spectrum of most 
X--ray pulsars is characterised by a relatively flat power law (photon index 
0.5--1.8), with a cut--off around 10--30~keV above which the spectrum is 
much steeper (White, Swank \& Holt 1983). This high energy cutoff is interpreted in 
terms of resonant cyclotron  absorption in the vicinity of the polar caps. In 
those X--ray pulsars in which cyclotron line features are observed, the 
cutoff energy, $E_{cut}$, was empirically determined to be related to the 
cyclotron line energy through $E_{cyc} \sim  (1.2-2.5) E_{cut}$ 
(Makishima et al. 1990, 
1992). Therefore, $E_{cut}$ can be used to approximately estimate the  
magnetic field strength (see however the case of EXO~2030+375; Reynolds, 
Parmar \& White 1993). 
With the exception of 4U~1626--67, the photon indeces of the spectra in 
our sample (measured shortwards of 10 keV) are higher than the range 
measured in other X--ray pulsars. It is therefore likely that the part of the 
spectrum above $E_{cut}$ is predominantly observed in the X--ray pulsars in 
our sample. Due to the combined effects of poor statistics and photoelectric 
absorption, a cutoff below 2--3~keV would remain undetected in the 
available spectra of 4U~0142+61, 1E~1048.1--5937 and 1E~2259+586. We 
conclude that for the  pulsars in our sample $E_{cut}$ $<$ 2--3~keV and 
therefore $B$ $<$ 8 $\times$ 10$^{11}$~G. 
\begin{figure}[bht]
\centerline{\psfig{figure=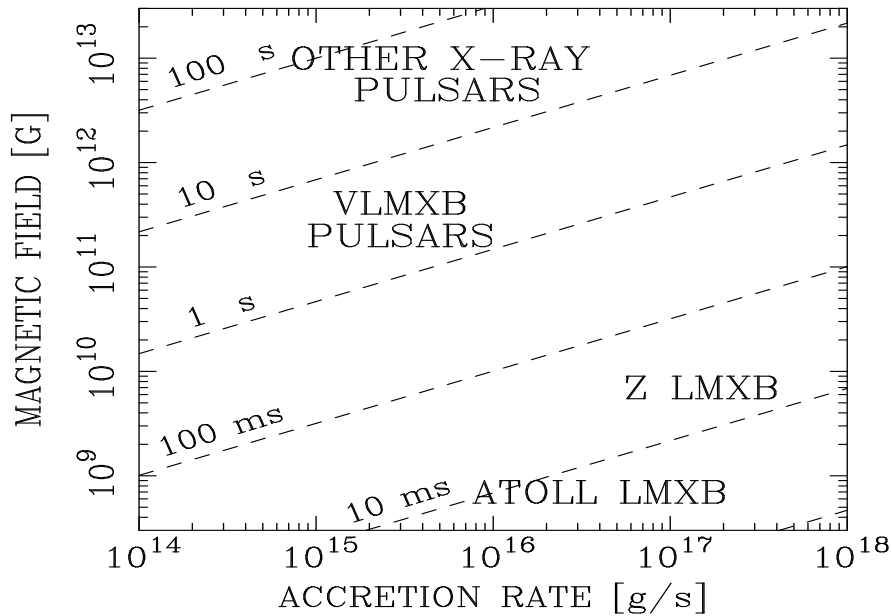,width=14cm,height=8.5cm} }
\caption[h]{Schematic representation of the position occupied by different 
classes of X--ray binaries in the $B$, \.{M} diagram. Lines of different 
equilibrium spin periods are also plotted.}
\end{figure}
These values are substantially lower than those measured (or 
inferred) from most other X--ray pulsars (10$^{12}-10^{13}$ G). Assuming $B = 
5$ $\times 10^{11}$ 
G (based on the analogy with 1E~2259+586),  Eq.~1 yields  for the sources 
in our sample  luminosities of 1--4 $\times$ 10$^{35}$ erg s$^{-1}$. The corresponding 
distances, $d_{eq}$, are reported in Table~1. These distance and luminosity 
values are generally in good agreement with the X--ray and optical 
observational data described in Section 2.  The only exception is 
4U~1626--67,  for which $d_{eq}\sim 1.7$ kpc is outside the range of 3--6~kpc 
derived by 
Levine et al. (1988) based on the spin--up observed before 1990. This 
indicates that 4U~1626--67 has a somewhat higher magnetic field ($d_{eq}$ 
scales linearly with $B$), as also suggested by its hard spectrum. A similar 
conclusion is reached if, in addition to Eq.~1, the equation for the accretion 
torque (cf. eq. 28 in Henrics 1983) is used together with  the period 
derivative observed during the spin--up phase (\.{P}/P$\sim$--2 $\times$10$^{-4}$ 
yr$^{-1}$) and a 
luminosity decrease of a factor $\sim$4 observed in correspondence to the \.{P} 
reversal (Bildsten et al. 1994).
This provides an independent rough 
estimate of $B\sim$1.1 $\times$ 10$^{12}$ G and $L_{eq} \sim 9 \times$ 10 
$^{35}$ erg s$^{-1}$. 

\section{Conclusions}

The most likely interpretation is that 
the X--ray pulsars 4U~0142+61, 1E~1048.1--5937, 4U~1626--67, 
RX~J1838.4--0301 and 1E~2259+589 are rotating close to their 
equilibrium spin periods and accreting from similar low mass companions. 
Their magnetic fields are likely of the order of a few 10$^{11}$ G, i.e. 
lower than in HMXB pulsars, but still much higher than  those of non--pulsating 
LMXBs (see Fig.~6). 

     Though the orbital period is (possibly) known only for 4U~1626--67, 
the  faintness of the optical counterpart and the absence of Doppler modulations in the X--ray pulses, indicate that these systems have
small  orbital periods and/or companions of very low mass. The main 
contribution to the optical emission is thus expected to come from the 
X--ray heated accretion disk. An empirical relation between absolute 
magnitude, X--ray luminosity and orbital period for LMXBs has   
been derived by van Paradijs \& McClintock (1994). For $L_x \sim 1-4 ~\times$ 
10$^{35}$ erg s$^{-1}$ 
it predicts $M_v \sim 5 - 1.5 Log(P_{orb}/1 $hr). Therefore, we would expect to 
detect the optical counterparts only for the closest and least absorbed of 
our LMXB pulsars, as indeed is the case (the optical absorption of 
4U~1626--67 is only A$_v < 0.3$; van Paradijs et al. 1986).
The relatively low luminosities imply that only the closest members of 
this class of LMXBs have so far been discovered. It is also possible that 
some of the already known, but poorly studied, LMXBs contain a pulsar 
with a spin period in the 5--10~s range. 

Future studies of this new class of X--ray pulsars should concentrate on: (a)
more accurate X--ray pulse timing studies that could reveal unambiguously the
presence of a companion star;  (b) deeper searches for the optical
counterparts; (c) periodicity searches in low 
luminosity X--ray sources close to the galactic plane. 

\acknowledgements
This work was partially supported through ASI grants. 
This research has made use of data obtained through the High Energy
Astrophysics Science Archive Research Center Online Service, provided by the
NASA-Goddard Space Flight Center. EXOSAT ME data were also extracted from the 
High Energy Astrophysics Database Service at the Brera Astronomical 
Observatory.


\begin{thebibliography}{}   

\bibitem[]{}  Angelini, L. et al.: 1995, \APJL 449 41.
\bibitem[]{} Baykal, A. \& Swank, J.H.: 1995, \APJ in~press.  
\bibitem[]{} Bildsten, L. et al.: 1994, AIP Conf. Proceedings 304 290.
\bibitem[]{} Carlini, A. \&  Treves, A.: 1989, \AAP 215 283.
\bibitem[]{} Coe, M.J. \&  Jones, L.R: 1992, \MN 259 191.
\bibitem[]{} Coe, M.J., Jones, L.R., \& Letho, H.:  1994, \MN in~press.
\bibitem[]{} Corbet, R.H.D. \& Day, C.S.R: 1990, \MN 243 553.
\bibitem[]{} Corbet, R.H.D, Smale, A.P., Ozaki, M., et al.: 1995 \APJ submitted.
\bibitem[]{} Davies, S.R. \& Coe, M.J.: 1991, \MN 249 313.
\bibitem[]{} Thompson, C. \& Duncan, R.C.: 1993, \APJ 408 194.
\bibitem[]{} Fahlman, G.G. \& Gregory, P.C.: 1981, \NAT 293 202.
\bibitem[]{} Fahlman, G.G. et al.: 1982, \APJL 261 1.
\bibitem[]{} Ghosh, P. \& Lamb, F.K.: 1979, \APJ 232 259.
\bibitem[]{} Gregory, P.C. \& Fahlman, G.G.: 1980, \NAT 287 805.
\bibitem[]{} Hellier, C.: 1994, \MN 271 L21.
\bibitem[]{} Henrichs, H.F.: 1983, in "Accretion Driven Stellar X--ray Sources", ed. 
W.H.G. Lewin  \& E.P.J. van den Heuvel (Cambridge University Press).
\bibitem[]{} Hughes, V.A., et al.: 1984, \APJ 283 147.
\bibitem[]{} Israel, G.L., Mereghetti, S., Stella L.: 1994, \APJL 433 25.
\bibitem[]{} Iwasawa, K., Koyama, K., \& Halpern, J.P.: 1992, \PASJ 44 9.
\bibitem[]{} Koyama, K. et al.: 1989, \PASJ 41 461.
\bibitem[]{} Leisawitz, D.,  Bash, F.N. \& Thaddeus, P.: 1989, \APJS 70 731.
\bibitem[]{} Levine, A. et al.: 1988, \APJ 327 732.
\bibitem[]{} Lutovinov, A.A., Grebenev, S.A., Sunyaev, R.A., \& Pavlinskii, M.N.: 1994, 
Astronomy Letters 20 538.
\bibitem[]{} Makishima, K., et al.: 1990, \PASJ 42 295.
\bibitem[]{} Makishima, K., et al.: 1992, in "Frontiers of X--ray Astronomy", ed. Y. Tanaka \& 
K. Koyama, (Univ. Academy Press) 23.
\bibitem[]{} McClintock, J.E. et al.: 1977, \NAT 270 320.
\bibitem[]{} Mereghetti, S.: 1996, \APJ 455 598.
\bibitem[]{} Mereghetti, S. \& Stella, L.: 1995, \APJL 442 17.
\bibitem[]{} Mereghetti, S., Caraveo, P. \& Bignami, G.F.: 1992, \AAP 263 172.
\bibitem[]{} Mereghetti, S., Stella, L. \&  De Nile, F.: 1993, \AAP 278 L23.
\bibitem[]{} Middleditch, J., Mason, K.O., Nelson, J.E., White, N.E.: 1981, \APJ 244 1001.
\bibitem[]{} Morini, M., Robba, N.R., Smith, A., \& van der Klis, M.: 1988, \APJ 333 777.
\bibitem[]{} Motch, C. et al.: 1991, \AAP 246 L24.
\bibitem[]{} Nagase, F.: 1989, \PASJ 41 1.
\bibitem[]{} Paczynski, B.: 1990, \APJL 365 9.
\bibitem[]{} Pravdo, S.H. et al.: 1979, \APJ 231 912.
\bibitem[]{} Rappaport, S.A. \& Joss, P.C.: 1983, in "Accretion Driven Stellar X--ray Sources", 
ed. W.H.G.Lewin \& E.P.J. van den Heuvel (Cambridge University Press).
\bibitem[]{} Rappaport, S. A. et al.: 1977, \APJL 217 29.
\bibitem[]{} Reynolds, A.P., Parmar, A.N., White, N.E.: 1993, \APJ 414 302.
\bibitem[]{} Schwentker, O.: 1994, \AAP 286 L47.
\bibitem[]{} Seward, F., Charles, P.A., Smale, A.P.: 1986, \APJ 305 814.
\bibitem[]{} Skinner, G.K. et al.: 1982, \NAT 297 568.
\bibitem[]{} Steinle, H., Pietsch, W., Gottwald, M. \& Graser, U.: 1987, \APSS 131 687.
\bibitem[]{} Usov, V.V.: 1994, \APJ 427 984.
\bibitem[]{} van Paradijs, J. \& McClintock, J.E.: 1994, \AAP 290 133.
\bibitem[]{} van Paradijs J., Taam R.E. \& van den Heuvel E.P.J. 1995, \AAP 299 L
41.
\bibitem[]{} van Paradijs, J., van Amerongen, S., Damen, E. \& van der Woerd, H.: 1986, \AAS 
63 71.
\bibitem[]{} White, N.E., et al.: 1987, \MN 226 645.
\bibitem[]{} White, N.E., Swank, J.H. \& Holt S.S.: 1983, \APJ 270 711.
\bibitem[]{} White, N.E., Stella, L., \& Parmar, A.N.: 1988, \APJ 324 363.
 
\end{thebibliography}
\end{document}